\documentclass[
    ,final            
    ,numberedheadings 
    ,cmfonts          
  ]
  {aipproc}
\usepackage{young}
\newcommand{\fr}{\frac}
\layoutstyle{6x9}
\def\1{\mbox{l\hspace{-0.53em}1}}

\begin{document}

\title{Group Theoretical Analysis of the Wave Function of the 
$[{\bf 70},1^-]$ Nonstrange Baryons in the $1/N_c$ Expansion}

\classification{12.39.--x, 11.15.Pg, 11.30.Hv}
\keywords      {Baryon Spectroscopy, $1/N_c$ Expansion, Group Theory}

\author{N. Matagne}{
  address={Institut f\"ur Theoretische Physik, Universit\"at Giessen, D-35392 Giessen, Germany}
}

\author{Fl. Stancu}{
  address={Universit\'e de Li\`ege, Institut de Physique B5, Sart Tilman, B-4000 Li\`ege 1, Belgium\\
E-mail: Nicolas.Matagne@theo.physik.uni-giessen.de, fstancu@ulg.ac.be}
}

\begin{abstract}
Using standard group theoretical techniques we construct the exact wave 
function of the  $[{\bf 70},1^-]$ multiplet in the orbital, spin and 
flavor space. This symmetric wave function is compared to that customarily
used in  the $1/N_c$ expansion, which is asymmetric.
The comparison is made by  analyzing the matrix elements of various
operators entering the mass formula. These matrix elements are calculated
by the help of isoscalar factors of the permutation group, specially 
derived for  this purpose as a function of $N_c$. 
We also compare two distinct methods used 
in the study of the  $[{\bf 70},1^-]$ multiplet. In the first method the
generators are divided into two parts, one part acting on a subsystem of
$N_c-1$ quarks called core and another  on the separated quark. 
In the second method  the system is treated as a whole. We show that the 
latter  is simpler and allows to 
clearly reveal the physically important operators in the mass 
formula. 

\end{abstract}

\maketitle

\section{Introduction}

For fifteen years the $1/N_c$ expansion of QCD, where $N_c$ is the number of 
colors \cite{tHo74,Wit79} has revealed itself to be an interesting and 
powerful approach for studying baryon spectroscopy. The method is based on an 
exact contracted  
SU$_c(2N_f)$ symmetry appearing in the large $N_c$ limit, $N_f$ being the 
number of flavors \cite{Gervais:1983wq,DM93}. For large $N_c$  this algebra  
becomes the SU($2N_f$) of the constituent quark model. 
Much work has been devoted to the ground state baryons where the 
operator reduction rules simplifies the expansion \cite{DJM94,DJM95}.
Usually higher order corrections of order $1/N^2_c$ are neglected.

For excited states the problem is more complicated. 
To include orbital excitations, by analogy to the
quark model, one can classify the large $N_c$ baryons 
according to an extended symmetry given by the 
direct product SU$(2N_f)\times$O(3). The group O(3) implies
the introduction of a spin-orbit and a tensor interaction.
It is a phenomenological fact that these
contributions are small so that the breaking of this symmetry is also small. 
An open problem is to investigate the validity 
of the $1/N_c$ expansion in this extended symmetry.

In the language of the quark model the excited states can be grouped 
into excitation bands $N$. In the $1/N_c$ expansion, the baryon masses
have been calculated for the lowest multiplets of all 
excited bands from $N = 1$ to $4$. In these bands the multiplets 
belong either to the symmetric  $[{\bf 56}]$  or to the 
mixed symmetric $[{\bf 70}]$ representation of SU(6).
The symmetry of the wave function of  excited baryons belonging to 
$[{\bf 56}]$ representation allows a similar treatment as that of
the ground state. The spin-flavor part being symmetric, the introduction of 
a symmetric orbital part does not modify the procedure.  In  the 
$[{\bf 70}]$ representation  (mainly the $[{\bf 70},1^-]$), the situation  
turned out to be more complicated. 
There is a standard scheme \cite{CCGL} where the wave function is
written as a product of a   written on spa6.log.
[stancu@ins
symmetric ground state core composed of $N_c-1$ quarks and an excited quark. 
In this approach, based on a Hartree picture, the $s^{N_c-1}p$ orbital part 
is not properly symmetrized and the excited quark is always the last quark. 
The flavor-spin part is also asymmetric and corresponds to a  
single term of the exact wave function. 

Recently, a new scheme, which avoids the separation into a core and an 
excited quark has been suggested \cite{MS1}. 
In that case the system is treated as a whole and the orbital-flavor-spin 
wave function is symmetric under any permutation of $N_c$ quarks. 
Some convincing quantitative arguments in favor of this procedure
can be found in Ref. \cite{MS2} 
The exact $[{\bf 70},1^-]$ wave function was written as a product of a  core
and a separated quark. Its orbital and the spin-flavor parts
are mixed symmetric such as to recover the exact symmetric orbital-spin-flavor 
wave function. The procedure is described in Sec. 4.1.

Using group theoretical arguments here we examine 
the relation between the exact  and the customarily used asymmetric wave
function. We argue that the description of the system  
is unsatisfactory
when the spin operator $S^2$ and the isospin $T^2$ operators are separated
into independent parts in terms of operators  acting separately
on the core and on the excited quark. Much better results are obtained
when we directly consider the operators $S^2$ and $T^2$ acting on
the whole system. In addition we examine the role of an operator 
constructed from the product of $S$, $T$ and $ G$ generators of SU(4).

\section{SU(4) generators as tensor operators}

The SU(4) generators $S_i$, $T_a$  and $G_{ia}$, 
globally denoted by $E_{ia}$ \ \cite{HP},
are components of an irreducible tensor operator
which transforms according to the adjoint representation $[211]$ of
dimension $\bf 15$ of SU(4).
We recall that the SU(4) algebra is
\begin{eqnarray}\label{ALGEBRASU4}
&[S_i,T_a] = 0,
~~~~~ [S_i,G_{ja}]  =  i \varepsilon_{ijk} G_{ka},\nonumber \\
&~~~~~ [T_a,G_{ib}]  =  i \varepsilon_{abc} G_{ic},\nonumber \\
&[S_i,S_j]  =  i \varepsilon_{ijk} S_k,
~~~~~ [T_a,T_b]  =  i \varepsilon_{abc} T_c,\nonumber \\
&[G_{ia},G_{jb}] = \fr{i}{4} \delta_{ij} \varepsilon_{abc} T_c
+\fr{i}{4} \delta_{ab}\varepsilon_{ijk}S_k.
\end{eqnarray}
As one can see, the tensor operators $E_{ia}$ 
are of three types:
$E_i$ ($i$ = 1,2,3) which form the subalgebra of SU(2)-spin, 
$E_a$ ($a$ = 1,2,3) which form the subalgebra of SU(2)-isospin
and  $E_{ia}$ which act both in the spin and the isospin spaces. They
are related to $S_i$, $T_a$ and $G_{ia}$ $(i=1,2,3;\ a=1,2,3)$ by
\begin{equation} \label{normes2}
E_i =\frac{S_i}{\sqrt{2}};~~~ E_a = \frac{T_a}{\sqrt{2}}; ~~~E_{ia} = \sqrt{2} G_{ia}.
\end{equation}

The matrix elements of every $E_{ia}$
between states belonging to the representation $[N_c-1,1]$ are given by
\begin{eqnarray}\label{GENsu4}
\lefteqn{\langle [N_c-1,1] I' I'_3 S' S'_3 | E_{ia} |
[N_c-1,1] I I_3 S S_3 \rangle  =   \sqrt{C^{[N_c-1,1]}(\mathrm{SU(4)})}   }
 \nonumber \\ & & \times
 \left(\begin{array}{cc|c}
	S   &    S^i   & S'   \\
	S_3  &   S^i_3   & S'_3
  \end{array}\right)
   \left(\begin{array}{cc|c}
	I   &   I^a   & I'   \\
	I_3 &   I^a_3   & I'_3
   \end{array}\right)
    \left(\begin{array}{cc||c}
	[N_c-1,1]    &  [211]   & [N_c-1,1]   \\
	S I  &   S^i I^a   &   S' I'
      \end{array}\right)_{\rho=1},
   \end{eqnarray}
where $C^{[N_c-1,1]}(\mathrm{SU(4)})=N_c(3N_c+4)/8$ is the eigenvalue of the 
SU(4) Casimir operator for the representation $[N_c-1,1]$. The three factors in the
second line are respectively an SU(2)-spin Clebsch-Gordan coefficient (CG), an SU(2)-isospin CG
and an SU(4) isoscalar factor.  The necessary iscocalar factors for the
derivation of the matrix elements of $E_{ia}$ have been calculated by 
Hecht and Pang \cite{HP}. Here, the phases and notations have been adapted
to our problem.

\section{The mass operator}

The mass operator $M$ is defined as a linear combination of
independent operators $O_i$ 
\begin{equation}
\label{massoperator}
M = \sum_{i} c_i O_i,
\end{equation} 
where the coefficients $c_i$ are reduced matrix elements that
encode the QCD dynamics and are  
determined from a fit to the existing data.
Here we are concerned with nonstrange baryons only.
The building blocks of the operators $O_i$ are the 
SU(2$N_f$) generators $S_i$, $T_a$ and $G_{ia}$ and the 
SO(3) generators $\ell_i$. 
Their general form is
\begin{equation}\label{OLFS}
O_i = \frac{1}{N^{n-1}_c} O^{(k)}_{\ell} \cdot O^{(k)}_{SF},
\end{equation}
where  $O^{(k)}_{\ell}$ is a $k$-rank tensor in SO(3) and  $O^{(k)}_{SF}$
a $k$-rank tensor in SU(2)-spin, but invariant in SU($N_f$).
Thus $O_i$ are rotational invariant.
For the ground state one has $k = 0$. The excited
states also require  $k = 1$ and $k = 2$ terms.

The spin-flavor (SF) operators  $O^{(k)}_{SF}$ are combinations 
of SU(2$N_f$) generators, the lower 
index $i$ in the left hand side of (\ref{OLFS}) representing a specific 
combination.
Each $\it n$-body operator  is multiplied by an explicit  factor of 
$1/{N^{n-1}_c}$ resulting from the power counting rules.
Some compensating $N_c$ factors may arise in the matrix elements when 
$O_i$ contains a coherent operator such as $G^{ia}$ or $T^a$.

\subsection{The symmetric core plus excited quark procedure}

So far, for the baryons belonging to the $[{\bf 70}, \ell]$ multiplet  
the general practice was to consider that they consist of one distinguishable
excited quark moving in the collective potential generated by
$N_c-1$ ground state quarks \cite{CCGL}, the latter subsystem being called
core. The wave function of the core is symmetric  both in the
orbital (O) and flavor-spin (FS) spaces, which makes the treatment
of the core analogous to that of ground state baryons.
This description is known as the Hartree picture. 

To proceed, one defines separate  SU($2N_f$) generators that act on the 
excited quark
$s^i$, $t^a$ and $g^{ia}$ and  $S^i_c$, $T^a_c$ and $G^{ia}_c$  
that act on the core. Thus one has 
\begin{equation}
\label{separation}  
S^i = s^i + S^i_c, ~~~ T^a = t^a + T^a_c, ~~~ G^{ia} =g^{ia} + G^{ia}_c.
\end{equation}
As a consequence, the number of linearly independent operators $O_i$ increases 
tremendously  and the number of coefficients $c_i$, 
to be determined, becomes much larger than the experimental data
available. For example, for the $[{\bf 70},1^-]$ multiplet with $N_f = 2$ 
one has 12 linearly independent
operators up to order $1/N_c$  included \cite{CCGL}. For example,
there is one operator of order $N^1_c$: $ N_c \ \1 $, 
three operators of order $N^0_c$: $\ell \cdot s$, $1/N_c\; \ell\cdot t\cdot G_c$,
$1/N_c\; \ell^{(2)} \cdot g \cdot G_c$ and  8 operators of order $N^{-1}_c$:
$1/N_c\; t\cdot T_c$, $1/N_c\; \ell \cdot S_c$, $1/N_c\; 
\ell\cdot g\cdot T_c$,  $1/N_c\; S^2_c$,
$1/N_c\; s\cdot S_c$, $1/N_c\; \ell^{(2)}\cdot s\cdot S_c$, $1/N^2_c\; \ell^{(2)}\cdot t\cdot \{S_c,G_c\}$
and $1/N^2_c\; \ell\cdot g\cdot \{S_c,G_c\}$.
Then, in making the 
fit to the data, one faces the difficult situation of selecting among them
the physically most dominant operators. 
We recall that there are only 7 nonstrange 
resonances belonging to this band. So one must select 7 out of
12  operators. 
Consequently, in selecting the operators one risks to make an
arbitrary choice  \cite{CCGL}. A much simpler method can be 
found, as shown below.

\subsection{A simpler procedure}
A simpler procedure is to avoid the splitting of the 
generators and the decoupling of the wave function and  to 
consider instead only the global generators $S^i$, $T^a$ and $G^{ia}$
acting on the whole system of $N_c$ quarks. However, the approach is not free 
of difficulties as the derivation of the matrix elements of the operators 
is more involved for a mixed symmetric wave function. Presently, the study of 
strange baryons is not possible. In the case of three flavors, 
one needs the analogue of Eq. (\ref{GENsu4}) containing the corresponding 
SU(6) isoscalar. These factors have not been calculated yet. 

In addition to the fact that it uses an exact wave function, this approach
implies only seven independent operators up to order 
$\mathcal{O}(1/N_c)$ appearing in 
the mass operator: the order $N_c$ operator $N_c\ \1$, three operators of 
order 1, $\ell\cdot s$, $1/N_c\; \ell^{(2)}\cdot G \cdot G$ 
and $1/N_c\; \ell \cdot T \cdot G$ and three operators of order 
$\mathcal{O}(1/N_c)$, namely $1/N_c\; S^2$, $1/N_c\; T^2$ and $1/N_c^2\; S\cdot T\cdot\ G$.

\section{The exact wave function}

The total wave function is the product of the orbital (O), the spin (S), 
the flavor (F) and the color (C) parts. The color part being always 
antisymmetric, in order to fulfill the Fermi statistics, 
the orbital-spin-flavor must be symmetric. As the mass operator does not 
involve color operators, the color being integrated out, we are concerned 
with the orbital-spin-flavor part only. 

Here, as we are interested in the multiplet $[{\bf 70},1^-]$,  the orbital 
and the spin-flavor parts must have both the mixed symmetry $[N_c-1,1]$. 
In terms of inner products of the permutation group $S_{N_c}$, the wave 
function takes the form
\begin{equation}
\label{EWF}
|[N_c]1 \rangle = \frac{1}{\sqrt{N_c-1}}
\sum_{Y} |[N_c-1,1] Y \rangle_{O}  |[N_c-1,1] Y \rangle_{FS},
\end{equation}
where $Y$ is the corresponding Young tableau. Here we sum over the $N_c-1$ 
possible standard Young tableaux. The factor $1/\sqrt{N_c-1}$ represents 
the CG coefficient of $S_{N_c}$ needed to construct a symmetric wave function $[N_c]$ from its mixed symmetric parts. 

\subsection{The decoupled wave function}
In order to calculate matrix elements of the operators listed in the Sec. 3.1
one must decouple the wave function as well.
Using the Racah's \emph{factorization lemma}, it is possible to decouple 
$N_c$th quark from the rest.  The $S_{N_c}$ CG coefficients can be factorized 
into an isoscalar factor times a CG coefficient of $S_{N_c-1}$. In the 
following, we need to know the position of the $N_c$th quark inside a given 
Young tableau. In that purpose, one introduces the integer $p$ which denotes the row where is 
the $N_c$th quark is located inside the Young tableau.

The  exact $[{\bf 70},1^-]$, but decoupled, wave function reads
\begin{eqnarray}
 \lefteqn{|\ell S J J_3; II_3\rangle =} \nonumber \\
& & \sum_{p, p', p'', \ell_c, \ell_q, m_\ell, m_q,\atop   m_c,m_s, m_1, m_2, i_1, i_2} a(p,\ell_c,\ell_q) \left(\begin{array}{cc|c}
	\ell_c    &  \ell_q   & \ell   \\
	m_c  &    m_q    & m_\ell 
      \end{array}\right) 
   \left(\begin{array}{cc|c}
	\ell    &    S   & J   \\
	m_\ell  &    m_s  & J_3 
      \end{array}\right) \nonumber \\
& & \times K([f']p'[f'']p''|[N_c-1,1]p) 
\left(\begin{array}{cc|c}
	S_c    &    \frac{1}{2}   & S   \\
	m_1  &         m_2        & m_s
      \end{array}\right)
 \left(\begin{array}{cc|c}
	I_c    &    \frac{1}{2}   & I   \\
	i_1    &       i_2        & I_3
      \end{array}\right) \nonumber \\
& & \times |\ell_c m_c\rangle  |S_cm_1\rangle|I_ci_1\rangle |\ell_q m_q\rangle |1/2m_2\rangle 
|1/2i_2\rangle,
\label{decex}
\end{eqnarray}
where $\ell_c$ and $\ell_q$ represent the angular momenta of the core and of 
the decoupled quark respectively and where $a(p,\ell_c,\ell_q)$ are the 
one-body fractional parentage coefficients to decouple the $N_c$th quark
from the rest in the orbital part. These are given by \cite{MS2}
\begin{eqnarray}
a(2,\ell_c =0, \ell_q = 1) & = & \sqrt{\frac{N_c-1}{N_c}}, \label{2coreground}\\
a(2,\ell_c =1, \ell_q = 0) & = & -\sqrt{\frac{1}{N_c}}, \label{2coreex} \\
 a(1,\ell_c =1, \ell_q = 0) & = & 1.
\end{eqnarray}
The isoscalar factors $K([f']p'[f'']p''|[N_c-1,1]p)$ used in 
Eq. (\ref{decex}) are given in Appendix A 
(Tables \ref{spin1/2n}, \ref{spin3/2n} and \ref{spin1/2delta}). The columns 
corresponding to $p=1$ have been derived in Ref. \cite{MS2}. If we compare 
Eq. (\ref{decex}) with Eq. (3.4) of Ref. \cite{CCGL} one can notice that 
in the latter only the terms with $p=2$ have been taken into account. 
Furthemore, as the core was assumed to be in the ground state, the authors  
had considered $a(2,\ell_c =1, \ell_q = 0) = 0$ and 
$a(2,\ell_c =0, \ell_q = 1) = 1$. Thus the wave function of Ref. \cite{CCGL}
breaks $S_{N_c}$ symmetry. As it represents only one part from the exact
wave function we shall call it approximate or asymmetric.

Tables \ref{spincomp} and \ref{isospincomp} show the matrix elements for 
some spin and the isospin operators respectively calculated with the exact 
and with the approximate wave function. One can notice that the analytic 
expressions are different. Consequently, one expects the $c_i$ coefficients 
determined from the fit to the data to be different if we use the exact 
or the approximate wave function. 

\begin{table}[h!]
\caption{Matrix elements of the spin operators calculated with the 
approximate (Ref. \cite{CCGL}) and the exact, Eq. (\ref{decex}), wave functions, with $s$ and $S_c$ defined by
Eq. (\ref{separation}).}
\label{spincomp}
\renewcommand{\arraystretch}{1.25}
\begin{tabular}{c|cc|cc} 
\hline \hline
 & \multicolumn{2}{c|}{$\langle s\cdot S_c\rangle$} & \multicolumn{2}{c}{$\langle S^2_c\rangle$}\\ 
& Approx. w.f. & Exact w.f. & Approx. w.f. & Exact w.f. \\ \hline
$^2 8$ & $ -\frac{N_c+3}{4N_c}$ & $-\frac{3(N_c-1)}{4N_c}$ & $\frac{N_c+3}{2N_c}$  & $\frac{3(N_c-1)}{2N_c}$ \\ 
$^4 8$ & $\frac{1}{2}$  & $-\frac{3(N_c-5)}{4N_c}$  & 2 & $\frac{3(3N_c-5)}{2N_c}$  \\ 
$^2 10$ &  $-1$ & $-\frac{3(N_c-1)}{4N_c}$& 2 &  $\frac{3(N_c-1)}{2N_c}$ \\ \hline \hline
\end{tabular}
\end{table}

\begin{table}[h!]
\caption{Matrix elements of the isospin operators calculated with the 
approximate (Ref. \cite{CCGL}) and the exact ,Eq. (\ref{decex}), wave functions, with $t$ and $T_c$ defined by
Eq. (\ref{separation}).}
\label{isospincomp}
\renewcommand{\arraystretch}{1.25}
\begin{tabular}{c|cc|cc} \hline \hline
 & \multicolumn{2}{c|}{$\langle t\cdot T_c\rangle$} & \multicolumn{2}{c}{$\langle T^2_c\rangle$}\\ 
& Approx. w.f. & Exact w.f. & Approx. w.f. & Exact w.f. \\ \hline
$^2 8$ & $ -\frac{N_c+3}{4N_c}$ & $-\frac{3(N_c-1)}{4N_c}$ & $\frac{N_c+3}{2N_c}$  & $\frac{3(N_c-1)}{2N_c}$ \\ 
$^4 8$ &  $-1$ & $-\frac{3(N_c-1)}{4N_c}$& 2 &  $\frac{3(N_c-1)}{2N_c}$ \\ 
$^2 10$ & $\frac{1}{2}$  & $-\frac{3(N_c-5)}{4N_c}$  & 2 & $\frac{3(3N_c-5)}{2N_c}$  \\ \hline\hline
\end{tabular}
\end{table}


\subsection{The global wave function}

As already mentioned above, one can write the exact $[{\bf 70},1^-]$ 
states without decoupling them  into a core and an excited  quark.  
If there is no decoupling, there is no need to
specify $Y$, the matrix elements being identical for all $Y$'s,
due to Weyl's duality between a linear group and a symmetric group
in a given tensor space\footnote{see Ref. \cite{book}, Sec 4.5.}.
Then the explicit form of a  wave function of total angular momentum 
$\vec{J} = \vec{\ell} + \vec{S}$ and  isospin $I$ is
\begin{eqnarray}\label{WF}
\lefteqn{|\ell SJJ_3; I I_3 \rangle  = } \nonumber \\
& & \sum_{m_\ell,m_s}
      \left(\begin{array}{cc|c}
	\ell    &    S   & J   \\
	m_\ell  &    m_s  & J_3 
      \end{array}\right) 
|[N_c-1,1]\ell m_{\ell} \rangle
|[N_c-1,1] S m_s I I_3 \rangle,
\end{eqnarray}
each term containing an SU(2) CG coefficient, an 
orbital part 
$|[N_c-1,1]\ell m_{\ell} \rangle$ an a spin-flavor
part $|[N_c-1,1] S m_s I I_3 \rangle$.

\section{Results}
Here we present the results obtained from different fits to the experimental data. In the fits, the seven nonstrange resonances have been taken into account:  $^2 N_{1/2}(1538 \pm 18)$, $^4 N_{1/2}(1660 \pm 20)$,
$^2 N_{3/2}(1523 \pm 8)$,  $^4 N_{3/2}(1700 \pm 50)$,
$^4 N_{5/2}(1678 \pm 8)$,  $^2 \Delta_{1/2}(1645 \pm 30)$ and
$^2 \Delta_{3/2}(1720 \pm 50)$. 

In a first stage, we describe the fits obtained when we use the exact 
decoupled wave function. Afterwards, the results obtained with the global 
wave function are presented. 

In each case, we follow the spirit of the  Hartree picture which leads to a 
one-body spin-orbit operator $\ell \cdot s$. Its matrix elements are naturally
of order $N_c^0$. 

\subsection{With the decoupled wave function}

Tables \ref{defit1}--\ref{defit4} show the four different fits considered. Each 
time, the results obtained with the exact decoupled wave function are compared 
to the ones obtained with the approximate wave function.

In Table \ref{defit1},  we decouple the spin and the isospin operators. 
The operator $1/N_c\; T_c\cdot T_c$ is not present because its
matrix elements are identical to of those $1/N_c\; S_c\cdot S_c$
for the approximate wave function 
(see Tables \ref{spincomp} and \ref{isospincomp}). 
This is apparently a practical advantage in the decoupling scheme but
it has considerable physical disadvantages.
One can notice that even 
if the $\chi^2_{\mathrm{dof}}$ is satisfactory, the fit is very bad. Indeed, 
the value of $c_1$ is under-evaluated with respect to the commonly found
value 
of around 500 MeV and the values of $c_3$ and $c_4$ are exceedingly 
large and of opposite 
signs, which suggest some compensation. 

The fits presented in Tables \ref{defit2} and \ref{defit3} seem better. Here 
the linear combinations $2s\cdot S_c +S_c\cdot S_c+3/4=S^2$  
or $2t\cdot T_c + T_c \cdot T_c +3/4 =T^2$ have been introduced. The 
coefficient  $c_1$ 
has recovered its common value and the coefficients $c'_3$ or $c'_5$ have 
reasonable sizes, being about 70 MeV smaller for the exact wave function 
than for the approximate one. The quark-core operators 
$1/N_c\; s\cdot S_c$ or $1/N_c\; t\cdot T_c$ are still problematic because 
the value of their respective coefficients are too high, of order 500 MeV.

The last fit shown in Table \ref{defit4} correct these problems. All the 
coefficients have their natural sizes.  This  shows the necessity to 
consider the 
isospin-isospin operator on the same footing as the spin-spin operator. 
The values obtained  with the exact wave function and the approximate 
one are identical in this case because the matrix elements of the operators 
considered are the same for the two wave functions. By construction, in
both cases they are eigenfunctions of the total spin and isospin 
operators.

\begin{table}[h!]
\caption{List of operators $O_i$ and coefficients $c_i$ obtained 
in the numerical fit 
to the 7 known experimental masses of the lowest negative parity
resonances (see text).  For the operators 
$O_3$, $O_4$ and $O_5$ we use the matrix elements from Tables 
\ref{spincomp} and \ref{isospincomp}.} \label{defit1}
\renewcommand{\arraystretch}{1.25}
 \begin{tabular}{lcc}
\hline \hline
$O_i$  &   \hspace{0cm}   $c_i$(MeV) with approx. w.f.   & \hspace{0cm} 
$c_i$(MeV) with  w.f. Eq. (\ref{decex}) \\ \hline
$O_1 = N_c \ \1 $    &   $  211 \pm 23$ &  $299 \pm 20$ \\
$O_2 = \ell^i s^i$  &  $3 \pm 15$  & $3 \pm 15$ \\
$O_3 = \frac{1}{N_c} s^iS_c^i $  & $-1486 \pm 141$ &  $-1096\pm 125$\\
$O_4 = \frac{1}{N_c} S_c^iS_c^i$ &  $1182\pm 74$ &  $1545 \pm 122$ \\
$O_5 = \frac{1}{N_c} t^aT_c^a$ &  $-1508\pm 149$  &  $417 \pm 79$ \\ 
\hline
$\chi_{\mathrm{dof}}^2$       & $1.56$ &  $1.56$      \\ \hline \hline 
\end{tabular}
\end{table}

\begin{table}[h!]\caption{Same as Table \ref{defit1} but for $O'_3$, which combines  $O_3$ and $O_4$
instead of using them separately. }
\label{defit2}
\renewcommand{\arraystretch}{1.25}
 \begin{tabular}{lcc}
\hline \hline
$O_i$  &   \hspace{0cm}   $c_i$(MeV) with approx. w.f. & \hspace{0cm}  
$c_i$(MeV) with w.f. Eq. (\ref{decex}) \\ \hline
$O_1 = N_c \ \1 $    &   $513 \pm 4$ & $519 \pm 5$ \\
$O_2 = \ell^i s^i$    &   $3 \pm 15$ &  $3 \pm 15$ \\
$O'_3 = \frac{1}{N_c}\left(2 s^iS_c^i+S_c^iS_c^i+\frac{3}{4}\right)$ &  $219\pm 19$  &  $150\pm 11$ \\
$O_5 = \frac{1}{N_c} t^aT_c^a$ &  $417\pm 80$ &  $417\pm 80$ \\
\hline
$\chi_{\mathrm{dof}}^2$       &    $1.04$ &  $1.04$      \\ \hline \hline 
\end{tabular}
\end{table}

\begin{table}[h!]
\caption{Same as Table \ref{defit2} but combining isospin operators instead 
of spin operators. }\label{defit3}
\renewcommand{\arraystretch}{1.25}
 \begin{tabular}{lcc}
\hline \hline
$O_i$   &   \hspace{0cm}  $c_i$(MeV) with approx. w.f. & \hspace{0cm}  
$c_i$(MeV) with w.f. Eq. (\ref{decex}) \\ \hline
$O_1 = N_c \ \1 $    &   $516 \pm 3$ &  $522 \pm 3$ \\
$O_2 = \ell^i s^i$    &   $3 \pm 15$ & $3 \pm 15$ \\
$O_3 = \frac{1}{N_c}s^iS_c^i$ &  $450 \pm 33$ &  $450 \pm 33$ \\
$O'_5 = \frac{1}{N_c}\left(2 t^aT_c^a+T_c^aT_c^a+\frac{3}{4}\right)$ &  $214\pm 28$ &  $139 \pm 27$ \\ 
\hline
$\chi_{\mathrm{dof}}^2$       &   $1.04$ &  $1.04$      \\ \hline \hline 
\end{tabular}
\end{table}

\begin{table}[h!]
\caption{Fit with global operators proportional to the SU(2)-spin 
and SU(2)-isospin Casimir operators acting on the whole system (see text). }\label{defit4}
\renewcommand{\arraystretch}{1.15}
 \begin{tabular}{lcc}
\hline \hline
$O_i$  &   \hspace{0cm}  $c_i$(MeV) with  approx. w.f. & \hspace{0cm} 
$c_i$(MeV) with w.f. Eq. (\ref{decex}) \\ \hline
$O_1 = N_c \ \1 $    &   $484 \pm 4$ &  $484 \pm 4$ \\
$O_2 = \ell^i s^i$    &   $3 \pm 15$ &  $3 \pm 15$\\
$O'_3 = \frac{1}{N_c}\left(2 s^iS_c^i+S_c^iS_c^i+\frac{3}{4}\right)$ & $150 \pm 11$ &  $150 \pm 11$\\
$O'_5 = \frac{1}{N_c}\left(2 t^aT_c^a+T_c^aT_c^a+\frac{3}{4}\right) $ & $139 \pm 27$ & $139 \pm 27$  \\ \hline
$\chi_{\mathrm{dof}}^2$       & $1.04$ &  $1.04$ \\ \hline \hline
\end{tabular}
\end{table}
\subsection{With the global wave function}

With the simplified procedure described in Sec. 3.2 we can analyse
the role of  every of the  seven independent operators introduced there. 
Table \ref{operators} shows six different fits to the experiment. 
The operators $O_5,\ O_6$ and $O_7$ are normalized to allow their coefficients 
$c_i$ to have a natural size.
As already emphasized, the Fits 1--4 indicate that the coefficients of $O_3$ 
and $O_4$ have similar values.

The partial contributions and the theoretical masses obtained from the Fits 1 
and 6 are presented in Tables \ref{MASSES1} and \ref{MASSES2} respectively. 
From Table \ref{MASSES1} one can notice that the isospin-isospin 
operator in $\Delta$  masses plays a comparable role to the spin-spin 
operator in $N^*$ resonances. This was impossible to observe in
the symmetric core + excited quark procedure where the
isospin-isospin operators were always ignored, for reasons 
explained above.  From Table \ref{MASSES2} one can see that
the operator $O_7$, never included before, is dominant in all resonances except $^2 N_{J}$. 
This is a new finding, to be algebraically 
understood.

The operators $O_5$ and $O_6$ do not seem  to play an important role because, 
in addition to the fact that their coefficients are small and have an error 
bar comparable to their central values, their removal from the fit does not 
deteriorate it too badly. This justifies the previous choice presented in 
Section 5.1 where $O_5$ and $O_6$ were neglected. Of course, Fit 4 is identical 
to the one shown in Table \ref{defit4}.

Table \ref{operators} does not include a fit with $O_3$, $O_4$ and $O_7$ 
together. In our calculations we found that the simultaneous 
presence of $O_3$, $O_4$ and $O_7$ leads to a 
$\chi^2_{\mathrm{dof}} \approx 2 $. In this case the coefficients 
$c_3$ and $c_4$ become a bit higher, 
of the order 270 MeV and
 $c_7$ becomes negative suggesting a possible compensation  
with the contributions of $O_3$ and $O_4$.
This suggests that, by construction, $O_7$  contains part of the contribution 
of the spin-spin and isospin-isospin interactions. As mentioned above, the role 
of $O_7$ needs more investigation.

\begin{table}[h!]
\caption{List of operators and the coefficients resulting from 
numerical fits using the global wave function. The values of $c_i$ are indicated under the headings Fit n,
in each case.}
\label{operators}
\renewcommand{\arraystretch}{2} 
\begin{tabular}{lrrrrrr}
\hline
\hline
Operator \hspace{0cm} &\hspace{0.0cm} Fit 1 (MeV) & \hspace{0cm} Fit 2 (MeV) & \hspace{0cm}Fit 3 (Mev) &\hspace{0cm} Fit 4 (MeV) &\hspace{0cm} Fit 5 (MeV)&\hspace{0cm} Fit 6 (MeV) \\
\hline

$O_1 = N_c \ \1 $                            & $481 \pm5$  & $482\pm5$ &  $484\pm4$ &  $484\pm4$ & $498\pm3$ & $495\pm3$ \\
$O_2 = \ell^i s^i$                	     & $-31 \pm26$ & $-20\pm23$ & $-12\pm20$ & $3\pm15$ & $38\pm34$ & $-30\pm 25$ \\
$O_3 = \frac{1}{N_c}S^iS^i$                  & $161\pm 16$ & $149\pm11$ & $163\pm16$ & $150\pm11$ & $156\pm16$ & \\
$O_4 = \frac{1}{N_c}T^aT^a$                  & $169\pm36$  & $170\pm36$ & $141\pm27$ & $139\pm27$ & \\
$O_5 = \frac{15}{N_c}\ell^{(2)ij}G^{ia}G^{ja}$    & $-29\pm31$&            & $-34\pm30$&        & $-34\pm31$ & $-32 \pm 29$\\
$O_6 = \frac{3}{N_c}\ell^iT^aG^{ia}$            & $32\pm26$ & $35\pm26$ &            &      & $-67\pm30$ & $28\pm 20 $\\
$O_7 =  \frac{3}{N_c^2} S^i T^a G^{ia}$ &  & & & & & $649 \pm 61$ \\ 
\hline
$\chi_{\mathrm{dof}}^2$                                    & $0.43$      & $0.68$ & $0.94$           & $1.04$ & $11.5$ & $0.24$ \\
\hline \hline
\end{tabular}
\end{table}

\begin{table}[h!]
\caption{The partial contribution and the total mass (MeV) predicted by the 
$1/N_c$ expansion using Fit 1 and the global wave function. 
The last two columns give  the empirically known masses,
name and status.}\label{MASSES1}
\renewcommand{\arraystretch}{2}
\begin{tabular}{lrrrrrrrrrr}\hline \hline
                    &      \multicolumn{6}{c}{Part. contrib. (MeV)}  & \hspace{0cm} Total (MeV)   & \hspace{0cm}  Exp. (MeV)\hspace{0cm}& &\hspace{0cm}  Name, status \hspace{.0cm} \\

\cline{2-7}
                    &   \hspace{0cm}   $c_1O_1$  & \hspace{0cm}  $c_2O_2$ & \hspace{0cm}$c_3O_3$ &\hspace{0cm}  $c_4O_4$ &\hspace{0cm}  $c_5O_5$ &\hspace{0cm} $c_6O_6$   &        \\
\hline
$^2N_{\frac{1}{2}}$        & 1444 & 10 & 40 & 42 & 0 & -8  &   $1529\pm 11$  & $1538\pm18$ & & $S_{11}(1535)$****  \\
$^4N_{\frac{1}{2}}$        & 1444 &  26 & 201& 42 & -31& -20 &   $1663\pm 20$  & $1660\pm20$ & & $S_{11}(1650)$**** \\
$^2N_{\frac{3}{2}}$        & 1444 & -5  & 40 & 42 & 0  &  4  &   $1525\pm 8$   & $1523\pm8$  & & $D_{13}(1520)$****\\
$^4N_{\frac{3}{2}}$        & 1444 & 10  & 201& 42 & 25 & -8 &   $1714\pm45$   & $1700\pm50$ & & $D_{13}(1700)$***\\
$^4N_{\frac{5}{2}}$        & 1444 & -16 & 201& 42  & -6 & 12 &   $1677\pm8$    & $1678\pm8$  & & $D_{15}(1675)$****\\
\hline
$^2\Delta_{\frac{1}{2}}$  &  1444 & -10  & 40 & 211 & 0  & -40   & $1645\pm30$  & $1645\pm30$ & & $S_{31}(1620)$**** \\
$^2\Delta_{\frac{3}{2}}$  &  1444 & 5  & 40 & 211 & 0  & 20   & $1720\pm50$  & $1720\pm50$ & & $D_{33}(1700)$**** \\ 
\hline \hline
\end{tabular}
\end{table}

\begin{table}[h!]
\caption{The partial contribution and the total mass (MeV) predicted by the 
$1/N_c$ expansion using Fit 6 and the global wave function. 
The last two columns give  the empirically known masses,
name and status.}\label{MASSES2}
\renewcommand{\arraystretch}{2}
\begin{tabular}{lrrrrrrrrr}\hline \hline
                    &      \multicolumn{5}{c}{Part. contrib. (MeV)}  & \hspace{0cm} Total (MeV)   & \hspace{0cm}  Exp. (MeV)\hspace{0cm}& &\hspace{0cm}  Name, status \hspace{.0cm} \\

\cline{2-6}
                    &   \hspace{0cm}   $c_1O_1$  & \hspace{0cm}  $c_2O_2$ & \hspace{0cm}$c_5O_5$ &\hspace{0cm}  $c_6O_6$ &\hspace{0cm}  $c_7O_7$  &        \\
\hline
$^2N_{\frac{1}{2}}$        & 1486 & 10  & 0   & -7  & 41 &   $1529\pm 11$  & $1538\pm18$ & & $S_{11}(1535)$****  \\
$^4N_{\frac{1}{2}}$        & 1486 &  25 & -33 & -18 & 203&    $1663\pm 20$  & $1660\pm20$ & & $S_{11}(1650)$**** \\
$^2N_{\frac{3}{2}}$        & 1486 & -5  & 0   & 4   & 41  &     $1525\pm 7$   & $1523\pm8$  & & $D_{13}(1520)$****\\
$^4N_{\frac{3}{2}}$        & 1486 & 10  & 26  & -7  & 203 &    $1718\pm41$   & $1700\pm50$ & & $D_{13}(1700)$***\\
$^4N_{\frac{5}{2}}$        & 1486 & -15 & 7   & 11  & 203 &   $1677\pm8$    & $1678\pm8$  & & $D_{15}(1675)$****\\
\hline
$^2\Delta_{\frac{1}{2}}$  &  1486 & -10 & 0  & -35  & 203     & $1643\pm29$  & $1645\pm30$ & & $S_{31}(1620)$**** \\
$^2\Delta_{\frac{3}{2}}$  &  1486 & 5   & 0  & 18   & 203    & $1711\pm24$  & $1720\pm50$ & & $D_{33}(1700)$**** \\ 
\hline \hline
\end{tabular}
\end{table}

\section{Conclusions}

In principle, both the core + excited quark (Sec. 3.1) or the global (Sec. 3.2)
procedures are legitimate as long as they are combined with adequately 
constructed wave functions. 
 The core + excited quark procedure was the first attempt to study excited
 states in the $1/N_c$ expansion and naturally it has been proposed to make 
 the problem tractable at that time, by reducing it to the knowledge of ground 
 state matrix elements of the operators from the mass formula. Presently it
 seems obsolete.  We have 
shown that the global procedure is much more advantageous. It involves
a smaller number of independent operators which allow to
clearly identify the physically dominant operators in the mass 
formula. Our conclusion is that the spin operator $1/N_c\; S \cdot S$ 
is dominant in $N^*$ resonances, that the isospin operator $1/N_c\; T \cdot T$
is equally important in $\Delta$ resonances and that  $1/N_c\; S \cdot T \cdot G$
plays about the same dominant role both in $N^*$ and $\Delta$ resonances except
for $^2 N_{J}$, the
contribution to the mass being of the order of 200 MeV in all cases. 
More work should be done algebraically in order to understand
the role of $1/N_c\; S \cdot T \cdot G$. 

Moreover, we 
found that all operators containing the O(3) generators bring only
small contributions to the mass, from 4 MeV to 42 MeV. This finding
is consistent with  the constituent quark model
assumptions about the feebleness of the spin-orbit 
and the smallness of the tensor interaction. 

We have shown that the separation core + quark procedure fails to emphasize 
the role of the isospin operator. This is due to the inherent structure
of the asymmetric ground state core + excited quark wave function \cite{CCGL} 
which leads to equal matrix
elements for $S^2_c$ and $T^2_c$. Then the remaining part of the
isospin interaction $1/N_c\; t \cdot T_c$ becomes exceedingly large
if included in the fit
(Table 3) and it is not surprising that in all previous studies 
(see Ref. \cite{MS2} for a review) it has been totally ignored.

In conclusion, the simple procedure we advocate here brings much 
more physical insight into the study of the nonstrange $[{\bf 70},1^-]$ 
baryons in the $1/N_c$ expansion. It is an urgent need to determine
the isoscalar factors of SU(6) for mixed symmetric representations 
$[N_c-1,1]$ in order to extend  Eq. (3) to SU(6) and apply it
to strange baryons.

\appendix

\section{Isoscalar factors}

Here we reproduce the isoscalar factors needed to construct the exact decoupled 
wave function (see Eq. ({\ref{decex})).
Detailed information can be found in Refs. \cite{MS2,book,ISOSC}

\begin{table}[h!]
\caption{Isoscalar factors $K([f']p'[f'']p''| [f]p)$ for
$S=I=1/2$, corresponding to $^{2}8$ when $N_c=3$. The second column gives
results for $p = 1$ and the third for $p = 2$.}\label{spin1/2n}
\renewcommand{\arraystretch}{1.8}
 \begin{tabular}{cccc}\hline \hline
$[f']p'[f'']p''$   & $[N_c-1,1]1$  &\hspace{3cm}& $[N_c-1,1]2$  \\ \hline
$\left[ \frac{N_c+1}{2},\frac{N_c-1}{2}\right ] 1\left\lbrack \frac{N_c+1}{2},\frac{N_c-1}{2}\right \rbrack 1$ &    0    &  & $-\sqrt{\frac{3(N_c-1)}{4N_c}}$ \\
$\left\lbrack \frac{N_c+1}{2},\frac{N_c-1}{2}\right \rbrack 2 \left\lbrack \frac{N_c+1}{2},\frac{N_c-1}{2}\right \rbrack 2$ & $\sqrt{\frac{N_c-3}{2(N_c-2)}}$ &\hspace{1cm} & $\sqrt{\frac{N_c+3}{4N_c}}$ \\
$\left\lbrack \frac{N_c+1}{2},\frac{N_c-1}{2}\right \rbrack 2 \left\lbrack \frac{N_c+1}{2},\frac{N_c-1}{2}\right \rbrack 1$ &   $-\frac{1}{2}\sqrt{\frac{N_c-1}{N_c-2}}$ & & 0 \\
$\left\lbrack \frac{N_c+1}{2},\frac{N_c-1}{2}\right \rbrack 1 \left\lbrack \frac{N_c+1}{2},\frac{N_c-1}{2}\right \rbrack 2$ &   $-\frac{1}{2}\sqrt{\frac{N_c-1}{N_c-2}}$ & & 0 \\
\hline \hline
 \end{tabular}
\end{table}

\begin{table}[h!]
\caption{Isoscalar factors $K([f']p'[f'']p''| [f]p)$ for
$ S=3/2,\ I=1/2$, corresponding to $^{4}8$ when $N_c=3$. The second column gives
results for $p = 1$ and the third for $p = 2$.}\label{spin3/2n}
\renewcommand{\arraystretch}{1.8}
 \begin{tabular}{cccc}\hline \hline
$[f']p'[f'']p''$   & $[N_c-1,1]1$  &\hspace{3cm}& $[N_c-1,1]2$  \\ \hline
$\left\lbrack \frac{N_c+3}{2},\frac{N_c-3}{2}\right \rbrack 1 \left\lbrack \frac{N_c+1}{2},\frac{N_c-1}{2}\right \rbrack 1$ & $\frac{1}{2}\sqrt{\frac{(N_c-1)(N_c+3)}{N_c(N_c-2)}}$ & \hspace{1cm} & 0 \\
$\left\lbrack \frac{N_c+3}{2},\frac{N_c-3}{2}\right \rbrack 2 \left\lbrack \frac{N_c+1}{2},\frac{N_c-1}{2}\right \rbrack 2$ & $\frac{1}{2}\sqrt{\frac{5(N_c-1)(N_c-3)}{2N_c(N_c-2)}}$ & \hspace{1cm} & 0 \\
$\left\lbrack \frac{N_c+3}{2},\frac{N_c-3}{2}\right \rbrack 1 \left\lbrack \frac{N_c+1}{2},\frac{N_c-1}{2}\right \rbrack 2$ & $\frac{1}{2}\sqrt{\frac{(N_c-3)(N_c+3)}{2N_c(N_c-2)}}$ & & 1 \\
 $\left\lbrack \frac{N_c+3}{2},\frac{N_c-3}{2}\right \rbrack 2 \left\lbrack \frac{N_c+1}{2},\frac{N_c-1}{2}\right \rbrack 1$ & 0 & & 0 \\
\hline \hline
 \end{tabular}
\end{table}

\begin{table}[h!]
\caption{Isoscalar factors $K([f']p'[f'']p''| [f]p)$ for
$S=1/2,\ I=3/2$, corresponding to $^{2}10$ when $N_c=3$. The second column gives
results for $p = 1$ and the third for $p = 2$.}\label{spin1/2delta}
\renewcommand{\arraystretch}{1.8}
 \begin{tabular}{cccc} \hline \hline
$[f']p'[f'']p''$   & $[N_c-1,1]1$  &\hspace{3cm}& $[N_c-1,1]2$  \\ \hline
$\left\lbrack \frac{N_c+1}{2},\frac{N_c-1}{2}\right \rbrack 1 \left\lbrack \frac{N_c+3}{2},\frac{N_c-3}{2}\right
\rbrack 1$& $\frac{1}{2}\sqrt{\frac{(N_c-1)(N_c+3)}{N_c(N_c-2)}}$ & \hspace{1cm} & 0 \\
$\left\lbrack \frac{N_c+1}{2},\frac{N_c-1}{2}\right \rbrack 2 \left\lbrack \frac{N_c+3}{2},\frac{N_c-3}{2}\right \rbrack 2$ & $\frac{1}{2}\sqrt{\frac{5(N_c-1)(N_c-3)}{2N_c(N_c-2)}}$ & \hspace{1cm} & 0 \\
$\left\lbrack \frac{N_c+1}{2},\frac{N_c-1}{2}\right \rbrack 2 \left\lbrack \frac{N_c+3}{2},\frac{N_c-3}{2}\right \rbrack 1$ & $\frac{1}{2}\sqrt{\frac{(N_c-3)(N_c+3)}{2N_c(N_c-2)}}$ & & 1 \\
$\left\lbrack \frac{N_c+1}{2},\frac{N_c-1}{2}\right \rbrack 1 \left\lbrack
\frac{N_c+3}{2},\frac{N_c-3}{2}\right \rbrack 2$ & 0 & & 0 \\
\hline \hline
 \end{tabular}
\end{table}



\end{document}